\documentclass[final,letterpaper,5p]{elsarticle}
\usepackage{lineno}
\usepackage{graphicx,color}
\usepackage{amssymb}
\usepackage{amsmath}
\usepackage{latexsym}
\usepackage{amsxtra}
\usepackage{amscd}
\usepackage{dcolumn}
\usepackage{bm}
\usepackage{hyperref}
\biboptions{numbers,sort&compress}
\graphicspath{{./}}
\let\oldsqrt\sqrt 
\def\sqrt{\mathpalette\DHLhksqrt}
\def\DHLhksqrt#1#2{\setbox0=\hbox{$#1\oldsqrt{#2\,}$}\dimen0=\ht0
\advance\dimen0-0.2\ht0
\setbox2=\hbox{\vrule height\ht0 depth -\dimen0}%
{\box0\lower0.4pt\box2}}

\newcommand{\nuc}[2]{$^{#1}$#2}
\journal{Physics Letters B}

\begin{document}
\begin{frontmatter}

\title{Shape coexistence and isospin symmetry in $A=70$ nuclei:\\ Spectroscopy of the $T_z =-1$ nucleus \nuc{70}{Kr}}

\author[ut,rnc]{K.~Wimmer} 
\author[cea]{W.~Korten} 
\author[gsi,jlu]{T.~Arici} 
\author[rnc]{P.~Doornenbal} 
\author[cce]{P.~Aguilera} 
\author[val,has]{A.~Algora}
\author[ut]{T.~Ando} 
\author[rnc]{H.~Baba} 
\author[bor]{B.~Blank} 
\author[pad]{A.~Boso}
\author[rnc]{S.~Chen} 
\author[cea]{A.~Corsi} 
\author[uy]{P.~Davies} 
\author[leg]{G.~de Angelis} 
\author[gan]{G.~de France} 
\author[cea]{D.~T.~Doherty} 
\author[gsi]{J.~Gerl} 
\author[tum]{R.~Gernh\"{a}user} 
\author[uy]{D.~Jenkins} 
\author[ut]{S.~Koyama} 
\author[rnc]{T.~Motobayashi} 
\author[ut]{S.~Nagamine} 
\author[ut]{M.~Niikura} 
\author[cea,rnc]{A.~Obertelli} 
\author[tum]{D.~Lubos} 
\author[val]{B.~Rubio} 
\author[uo]{E.~Sahin} 
\author[ut]{T.~Y.~Saito} 
\author[ut,rnc]{H.~Sakurai} 
\author[uy]{L.~Sinclair} 
\author[rnc]{D.~Steppenbeck} 
\author[ut]{R.~Taniuchi} 
\author[uy]{R.~Wadsworth} 
\author[cea]{M.~Zielinska}

\address[ut]{Department of Physics, The University of Tokyo, 7-3-1 Hongo, Bunkyo-ku, Tokyo 113-0033, Japan}
\address[rnc]{RIKEN Nishina Center, 2-1 Hirosawa, Wako, Saitama 351-0198, Japan}
\address[cea]{IRFU, CEA, Universit\'{e} Paris-Saclay, F-91191 Gif-sur-Yvette, France}
\address[gsi]{GSI Helmholtzzentrum f\"{u}r Schwerionenforschung, D-64291 Darmstadt, Germany}
\address[jlu]{Justus-Liebig-Universit\"{a}t Giessen, D-35392 Giessen, Germany}
\address[cce]{Comisi\'{o}n Chilena de Energ\'{i}a Nuclear, Casilla 188-D, Santiago, Chile}
\address[val]{Instituto de Fisica Corpuscular, CSIC-Universidad de Valencia, E-46071 Valencia, Spain}
\address[has]{Institute of Nuclear Research of the Hungarian Academy of Sciences, Debrecen H-4026, Hungary}
\address[bor]{CENBG, CNRS/IN2P3, Universit\'{e} de Bordeaux F-33175 Gradignan, France}
\address[pad]{Istituto Nazionale di Fisica Nucleare, Sezione di Padova, I-35131 Padova, Italy}
\address[uy]{Department of Physics, University of York, YO10 5DD York, United Kingdom}
\address[leg]{Istituto Nazionale di Fisica Nucleare, Laboratori Nazionali di Legnaro, I-35020 Legnaro, Italy}
\address[gan]{GANIL, CEA/DSM-CNRS/IN2P3, F-14076 Caen Cedex 05, France}
\address[tum]{Physik Department, Technische Universit\"{a}t M\"{u}nchen, D-85748 Garching, Germany}
\address[uo]{Department of Physics, University of Oslo, PO Box 1048 Blindern, N-0316 Oslo, Norway}

\begin{abstract}
Excited states in the $T_z=-1$ nucleus \nuc{70}{Kr} have been populated using inelastic scattering of a radioactive \nuc{70}{Kr} beam as well as one- and two-neutron removal reactions from \nuc{71,72}{Kr} at intermediate beam energies. The level scheme of \nuc{70}{Kr} was constructed from the observed $\gamma$-ray transitions and coincidences. Tentative spin and parity assignments were made based on comparison with the mirror nucleus \nuc{70}{Se}. A second $2^+$ state and a candidate for the corresponding $4^+_2$ state suggest shape coexistence in \nuc{70}{Kr}.
\end{abstract}

\date{\today}
\end{frontmatter}

Shortly after the discovery of the neutron Heisenberg introduced isospin as a new symmetry~\cite{heisenberg32} in the nucleus. In this formalism, protons and neutrons are almost identical and regarded as nucleons with isospin quantum numbers $t_z=\pm 1/2$. Under the assumption of charge independence of the strong interaction, hence in-variance under rotation in the isospin space, the excitation energy spectra of mirror nuclei should be identical. In these nuclei, which differ by the interchange of proton and neutron number, differences arise from Coulomb effects and (weaker) isospin non-conserving terms in the nuclear interaction. The mirror energy differences (MED) between analog states in $T=1/2$ or 1 mirror pairs and triplet energy differences (TED) of $T=1$ triplets can therefore be used to obtain information on these isospin non-conserving interactions in nuclei (see~\cite{bentley07} for a review on the $f_{7/2}$ shell). The $A=70$ isotopes play an important role in this respect, as previous investigations of the Coulomb energy difference (CED) between $T_z=0$ and $T_z=1$ nuclei revealed that the $A=70$ nuclei show a different (negative) trend~\cite{deangelis01} than all other cases~\cite{narasingh07} studied so far in the $pf$ shell. A possible explanation for this unexpected behavior is related to the rapid evolution of nuclear shapes in these nuclei. 
In this scenario the shape of the ground state may differ between the mirror nuclei, e.g.~\nuc{70}{Kr} and \nuc{70}{Se}. Such a scenario is supported by the observed shape isomers \cite{chandler97, bouchez03} and the shape evolution in the chain of the Kr isotopes: While the less neutron-deficient Kr isotopes (above $A=74$) exhibit shape coexistence with a predominantly prolate ground state (with an excited oblate configuration), the most neutron-deficient isotopes (below $A=74$) are expected to have an oblate ground state. In \nuc{74}{Kr} both configurations are degenerate leading to strongly mixed $0^+$ states~\cite{poirier04}. For the isotopes with $A\ge74$ this complex scenario has been confirmed experimentally \cite{clement07, goergen05}, while experiments on \nuc{72}{Kr} have shown preliminary evidence for an oblate ground state deformation \cite{gade05,iwasaki14,briz15}. 
For \nuc{70}{Kr} various calculations predict an oblate~\cite{rodriguez14,moeller16} ground state shape also. The oblate shape of the $T_z=+1$ mirror nucleus \nuc{70}{Se} was inferred from the results of a combination of Coulomb excitation and lifetime measurement experiments~\cite{hurst07,ljungvall08}. Assuming isospin symmetry, the same shape would be expected for \nuc{70}{Kr}, while some theoretical calculations predict prolate dominated deformation in the ground state band~\cite{petrovici15}.
Recently, two transitions have been tentatively assigned to \nuc{70}{Kr}~\cite{debenham16}. Shell model calculations with isospin non-conserving interactions~\cite{kaneko12} reproduce the tentative $2^+$ and $4^+$ states without invoking a shape change between \nuc{70}{Se} and \nuc{70}{Kr}.

In this Letter we present extended spectroscopy of \nuc{70}{Kr}, the heaviest $T_z=-1$ nucleus accessible for detailed spectroscopic studies off the yrast line so far. Analogue reactions to \nuc{70}{Kr}, and its mirror nucleus \nuc{70}{Se}, have been performed with uniquely identified reaction products. Such direct mirrored reactions enable conclusions to be drawn about the structure of the states involved, as have been done in the past in, for example, the $fp$ shell~\cite{milne16}. Comparison of the relative final state exclusive cross sections of these reactions allow for tentative spin and parity assignments of states in \nuc{70}{Kr}. The observation of the $2^+_2$ state and a candidate for the $4^+_2$ state suggest the presence of shape coexistence in \nuc{70}{Kr}.

\begin{figure}[h]
\centering
\includegraphics[width=\columnwidth]{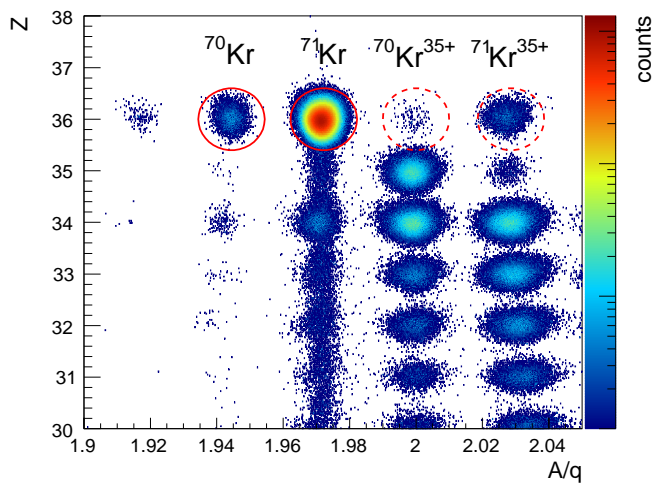}
\caption{Particle identification plot for the ZeroDegree spectrometer for \nuc{71}{Kr} impinging on the \nuc{9}{Be} secondary reaction target. Scattered \nuc{71}{Kr} as well as \nuc{70}{Kr} produced by neutron removal can be clearly separated from other reaction products. Hydrogen-like $q=35+$ charge states of the Kr isotopes appear at larger $A/q$ values.}
\label{fig:pid}
\end{figure}

The experiment was performed at the Radioactive Isotope Beam Facility operated by the RIKEN Nishina Center and CNS, University of Tokyo. Radioactive beams of \nuc{70,71,72}{Kr} were produced in projectile fragmentation of \nuc{78}{Kr} accelerated to 345 MeV/u impinging on a 5~mm thick Be target. The reaction fragments were separated in the first stage of the BigRIPS separator~\cite{kubo12} by means of their magnetic rigidity $B\rho$ and the energy loss in an Al degrader. Unambiguous identification was achieved in the second stage of BigRIPS through measurements of the time-of-flight, trajectory, and energy loss of the ions using the standard detection systems consisting of plastic scintillators, parallel plate avalanche counters and an ionization chamber. Data were taken in two settings, one centered on \nuc{72}{Kr}, one centered on \nuc{70}{Kr} while \nuc{71}{Kr} was also transmitted. Typical intensities and purities amounted to 6000, 250, and 15 particles per second and 64, 6.4, and 0.4\% for \nuc{72}{Kr}, \nuc{71}{Kr}, and \nuc{70}{Kr} in the respective settings. The beams then impinged on a 703(7)~mg/cm$^2$ thick secondary \nuc{9}{Be} reaction target. The average energy in the middle of the target was $\approx 140$~MeV/u. The target was surrounded by the DALI2 array, consisting of 186 NaI(Tl) crystals for highly efficient $\gamma$-ray detection~\cite{takeuchi14}. Standard $\gamma$ calibration sources were used for energy calibration, while the fields of nearby magnets were set to the value of the corresponding reaction setting. Second order polynomials were used in the calibration of light-output to the deposited $\gamma$-ray energy. Reaction products were identified by the ZeroDegree spectrometer~\cite{kubo12} through measurements of time-of-flight, $B\rho$, and $\Delta E$ in the same way as for BigRIPS. The particle identification in the ZeroDegree spectrometer is shown in Fig.~\ref{fig:pid} for the case of incident \nuc{71}{Kr} beam.
States in \nuc{70}{Kr} have been populated using three different reactions: (i) inelastic scattering of \nuc{70}{Kr} itself, (ii) one-neutron removal from \nuc{71}{Kr}, and (iii) two-neutron removal from \nuc{72}{Kr}. In addition, the analogue one-proton removal reaction from \nuc{71}{Br} to \nuc{70}{Se} was studied in the same setting. 

\begin{figure}[h!]
\centering
\includegraphics[width=\columnwidth]{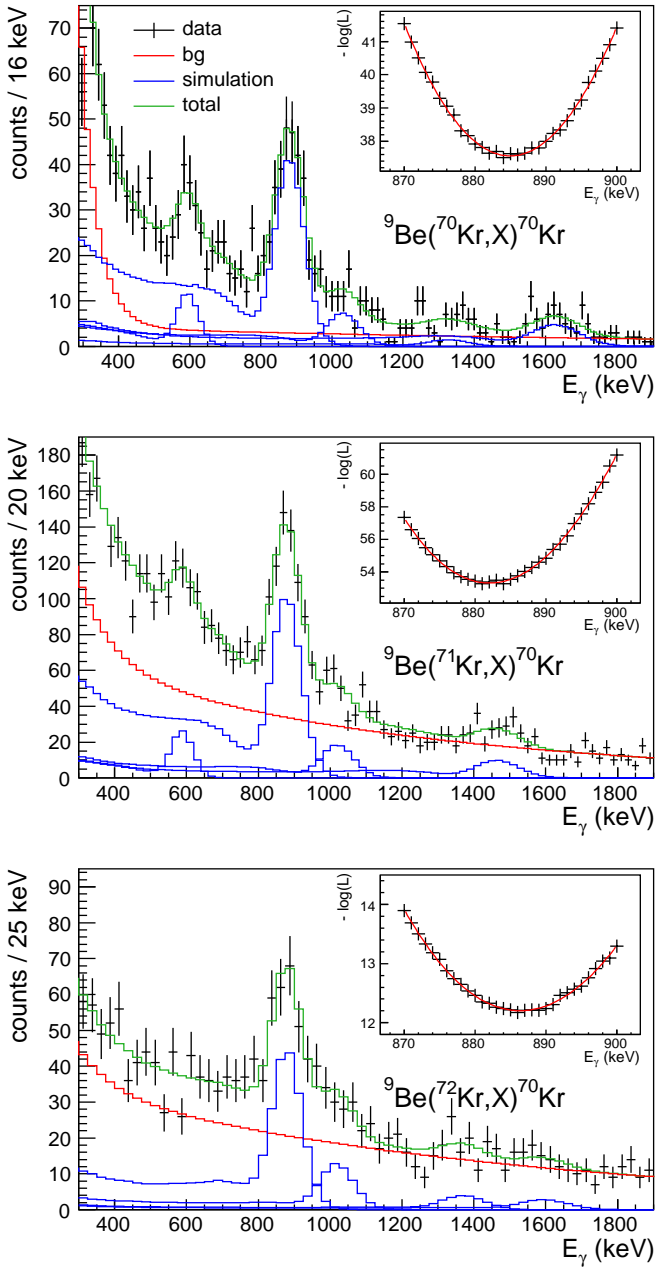}
\caption{Doppler corrected $\gamma$-ray energy spectra for \nuc{70}{Kr} populated in three different reactions. The measured $\gamma$-ray energies have been corrected for the Doppler-shift using the position of DALI2 crystals with respect to the target. The data are superimposed with the result of a likelihood fit of a continuous background and GEANT4 simulations of the DALI2 response function. The insets show the likelihood as a function of the simulated energy for the $2_1^+ \rightarrow 0_1^+$ transition.}
\label{fig:gamma}
\end{figure}

The Doppler corrected $\gamma$-ray energy spectra measured in coincidence with \nuc{70}{Kr} identified in the ZeroDegree spectrometer are shown in Fig.~\ref{fig:gamma} for the three different reactions (i) $-$ (iii).
Add-back has not been used to determine the transition energies since it produced a shift of the peak to slightly lower energies due to the non-linearity of the light production in the NaI crystals. For background reduction the hit multiplicity of DALI2 detectors was restricted to less than five. The data shown in Fig.~\ref{fig:gamma} are fitted with a combination of a continuous background and simulated response functions of the DALI2 array. For the determination of the transition energy, the simulated $\gamma$-ray energy has been varied and a likelihood fit has been performed to find the best matching energy. Considering only statistical errors, the energy of the $2^+$ state in \nuc{70}{Kr} amounts to 885(6), 882(5), and 886(11)~keV for the inelastic scattering, the one-neutron removal from \nuc{71}{Kr}, and the two-neutron removal reactions from \nuc{72}{Kr} cases, respectively.
The same fit procedure has also been applied to other nuclei produced in the same experiment under identical conditions. The well-known states in \nuc{68}{Se}, \nuc{70}{Se}, \nuc{70}{Br}, and \nuc{72}{Kr}, including their lifetimes, could be reproduced within the statistical error ($<5$~keV). The main source of systematic errors was the position of the DALI2 crystals with respect to the secondary reaction target due to the strong angular dependence of the Doppler shift effect at relativistic velocities. A shift of 2~mm in beam direction results in a difference of $\approx 4$~keV at 1~MeV. This shift would, however, be identical for all measured cases and can therefore be excluded as a source of systematic uncertainty. Another source of uncertainty is the energy loss in the target and the associated velocity of the ejectile used in the Doppler correction. The last source of systematic uncertainties considered here is the calibration of the light output of DALI2 and a possible variation of the gain over the course of the experiment.   
A weighted average of $884\pm 4 (\text{stat.}) \pm 5 (\text{sys.})$~keV for the excitation energy of the $2^+_1$ state is adopted in the following. Four other transitions have also been observed in the \nuc{9}{Be}(\nuc{71}{Kr},X)\nuc{70}{Kr} and \nuc{9}{Be}(\nuc{72}{Kr},X)\nuc{70}{Kr} neutron removal reactions and the inelastic scattering and their energies have been obtained from a similar likelihood fit. In all these cases the lifetime of the states were neglected. Due to the thick target, lifetimes up to 10~ps have no effect on the peak position ($<1$~keV), and even for 25~ps the shift is still within the statistical uncertainty quoted above ($\approx 5$~keV).

\begin{figure}[h]
\centering
\includegraphics[width=\columnwidth]{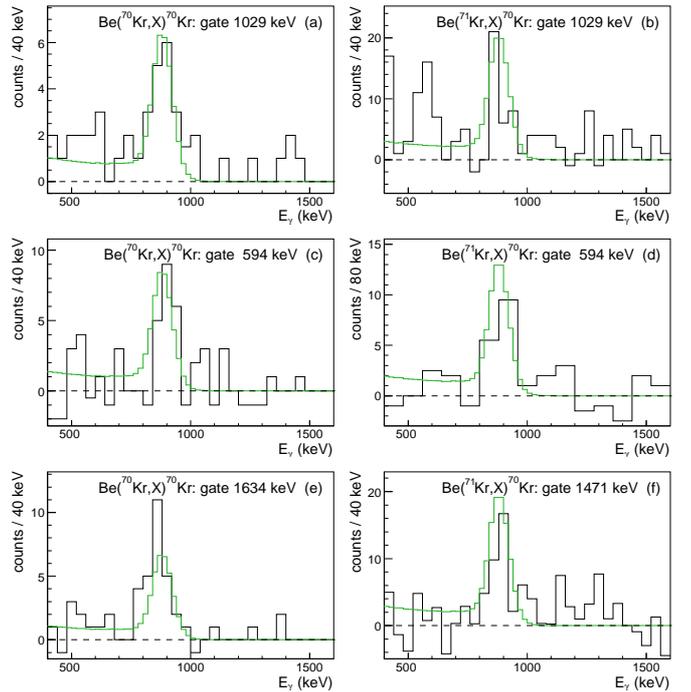}
\caption{Background subtracted, gated $\gamma$-ray energy spectra for \nuc{70}{Kr} populated in inelastic scattering (left column) and one-neutron removal (right column). The green curves show simulated response functions scaled for the number of expected coincidences based on the proposed level scheme. The simulations are shown with a binning of 20~keV/bin. Panels (a) and (b) are spectra gated on the 1029~keV transition, (c) and (d) on the 594~keV transition, (e) on the 1634~keV transition, and (f) on the 1471~keV transition which were observed only in the inelastic scattering or knockout reaction channel, respectively.}
\label{fig:coinc}
\end{figure}

The 884~keV line is the strongest transition observed in all three spectra, the $2^+_1$ state is therefore placed at 884(4)(5)~keV. 
In order to build the level scheme of \nuc{70}{Kr} $\gamma-\gamma$ coincidences have been analyzed. For the two-neutron removal reaction the obtained statistics were insufficient to perform a $\gamma-\gamma$ coincidence analysis. Gated $\gamma$-ray energy spectra for the other two reaction channels are shown in Fig.~\ref{fig:coinc}.
The transition at 1029(14)(5)~keV was observed in all three reaction channels. This transition was in coincidence with the decay of the $2^+_1$ state (Fig.~\ref{fig:coinc} (a) and (b)). Based on this and energy systematics along the Kr isotopic chain and the $A=70$ nuclei, the 1029~keV transition is assigned to the $4^+_1\rightarrow2^+_1$ transition, placing the first $4^+$ state at 1913(14)~keV.
\begin{figure*}[ht!]
\centering
\includegraphics[width=\textwidth]{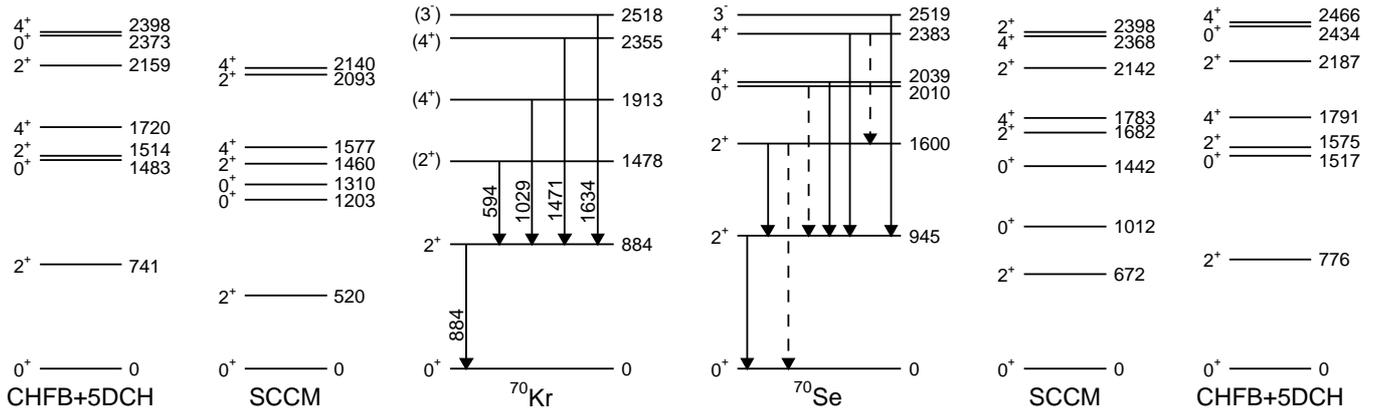}
\caption{Level scheme of \nuc{70}{Kr} and its mirror nucleus \nuc{70}{Se}. Spin and parity assignments for \nuc{70}{Kr} have been made based on this comparison and the observed decay pattern of states. For \nuc{70}{Se}, besides the $0^+_2$ state, only states which have been observed in the present work through the proton removal reaction from \nuc{71}{Br} are shown for clarity. Transitions indicated by dashed lines have not been clearly observed in this work. In \nuc{70}{Se} they are known to exist, however, the branching ratio is below the sensitivity limit of the present experiment. Theoretical level schemes have been calculated in the symmetry conserving configuration mixing (SCCM) approach~\cite{rodriguez14} and Hartree-Fock-Bogoliubov calculations mapped on a five-dimensional collective Hamiltonian (CHFB-5DCH)~\cite{delaroche10}.}
\label{fig:level}
\end{figure*}
This is at variance with the result obtained in a fusion reaction experiment~\cite{debenham16}, where the $2^+_1$ and $4^+_1$ states were tentatively observed at 870 and 1867~keV. In the fusion evaporation experiment, however, there was no particle identification possible and the statistical significance of the second transition was marginal.

The transition at 594(10)(5)~keV observed in the inelastic scattering of \nuc{70}{Kr} and the one-neutron knockout from \nuc{71}{Kr} was in coincidence with the 884~keV $2^+_1 \rightarrow 0^+_\text{gs}$ transition (Fig.~\ref{fig:coinc} (c) and (d)). A transition at 1471(25)(7) keV was observed only in the one-neutron removal reaction. Within error this transition energy matches the energy sum of 594 and 884~keV. This would suggest a state which decays both to the ground and the first $2^+$ states. However, the 594~keV transition is also observed in the inelastic scattering of \nuc{70}{Kr} on Be, while the 1471~keV transition is absent. Moreover, the 1471~keV transition is observed in coincidence with the 884~keV transition (Fig.~\ref{fig:coinc} (f)). Based on systematics and comparison with the mirror nucleus \nuc{70}{Se} (Fig.~\ref{fig:level}) the state at 1478(11)~keV is assigned $J^\pi = 2^+_2$. In \nuc{70}{Se} this state, located at 1600~keV, decays mainly to the $2^+_1$ state but also has a 25\% branch to the ground state. The 1471~keV transition is assigned to originate from a $4^+_2$ candidate at 2355(25)~keV. In the case of the inelastic scattering of \nuc{70}{Kr} a transition at 1634(24)(7)~keV is quite strongly populated and in coincidence with the 884~keV transition (Fig.~\ref{fig:coinc} (e)). Therefore, a state at 2518(24)~keV is tentatively assigned as $(3^-)$. Octupole states are frequently observed in inelastic scattering, while one would not expect a strong population of such a collective state in a single-particle knockout reaction.
Lastly, the $\gamma$-ray transitions observed at $1340^{+65}_{-99}$ and $1366^{+82}_{-53}$~keV in the inelastic scattering and two-neutron removal, respectively, are assumed to be the same. The obtained statistics are insufficient to analyze coincidences. 
\begin{table*}[ht]                 
  \caption{Relative intensities $I$, exclusive, and inclusive cross sections for the three reactions populating states in \nuc{70}{Kr}. The intensities are normalized to the 884~keV $2^+_1 \rightarrow 0^+_1$ transition. The cross sections are inclusive with respect to the potential isomeric contamination of the secondary beams.}
  \begin{center}
    \label{tab:intensities}
    {\renewcommand{\arraystretch}{1.1}\footnotesize
      \begin{tabular}{rrr|rr|rr|rr|rrr|rr} 
        \hline
        \multicolumn{3}{c}{ \nuc{70}{Kr} }\vline & \multicolumn{2}{c}{\nuc{9}{Be}(\nuc{70}{Kr},X)\nuc{70}{Kr}} \vline & \multicolumn{2}{c}{\nuc{9}{Be}(\nuc{71}{Kr},X)\nuc{70}{Kr}} \vline & \multicolumn{2}{c}{\nuc{9}{Be}(\nuc{72}{Kr},X)\nuc{70}{Kr}}\vline & \multicolumn{3}{c}{ \nuc{70}{Se} }\vline & \multicolumn{2}{c}{\nuc{9}{Be}(\nuc{71}{Br},X)\nuc{70}{Se}}\\
        $E$ (keV) & $J^\pi$ & $E_\gamma$ (keV) & $I_\gamma$ (\%) & $\sigma$ (mb) & $I_\gamma$ (\%) & $\sigma$ (mb) & $I_\gamma$ (\%) & $\sigma$ (mb) & $E$ (keV) & $J^\pi$ & $E_\gamma$ (keV) & $I_\gamma$ (\%) & $\sigma$ (mb) \\
        \hline
        0         & $0^+_1$ &             &        &         &        & 19.2(17) &         & 0.38(3)  & 0    & $0^+_1$ &      &        & 58(15) \\
        884(4)    & $2^+_1$ &  884(4)(5)  & 100(8) & 14(5)   & 100(6) & 2.3(6)   & 100(12) & 0.04(4)  & 945  & $2^+_1$ &  945 & 100(2) &  8(3)  \\
        1478(11)  & $2^+_2$ &  594(10)(5) &  16(5) & 6.0(20) &  16(4) & 0.8(2)   &         &          & 1600 & $2^+_2$ &  655 &  10(2) &  4(2)  \\
        1913(14)  & $4^+_1$ & 1029(14)(5) &  20(5) & 7.3(20) &  20(5) & 1.1(7)   & 35(10)  & 0.08(2)  & 2039 & $4^+_1$ & 1094 &  43(2) & 18(1)  \\
        2355(25)  & $4^+_2$ & 1471(25)(7) &        &         &  21(5) & 1.1(7)   &         &          & 2383 & $4^+_2$ & 1438 &   9(2) &  4(1) \\
        2518(24)  & $3_1^-$ & 1634(24)(7) &  25(6) & 9(2)    &        &          & 20(9)   & 0.04(2)  & 2519 & $3^-_1$ & 1574 &  11(5) &  5(2)  \\
                  &         &             &        &         &        &          &         &          & 2553 & $4^+_3$ & 1610 &   7(4) &  3(1)\\
        \multicolumn{2}{r}{unplaced}                    & 1353(50)(7) &   3(3) & 1(1)    &        &          & 24(10)  & 0.05(2)  &      &         &      &        &  \\
        \hline
        \multicolumn{3}{c}{inclusive}\vline &     &         &        & 24.5(12) &         & 0.59(7)   &      &         &      &        & 100(12) \\
        \hline    
      \end{tabular}
    }
  \end{center}
\end{table*}

The proposed level scheme of \nuc{70}{Kr} is shown in Fig.~\ref{fig:level} and compared with its mirror nucleus \nuc{70}{Se}. The mirror energy differences, defined as $\text{MED}(J^\pi_i) = E(J^\pi_i,T_z=-1) - E(J^\pi_i,T_z=+1)$ amount to $\text{MED}(2^+_1) = -61(4)$~keV and $\text{MED}(4^+_1) = -126(14)$~keV, smaller than the experimental values of Ref.~\cite{debenham16} but reproduced by shell model calculations using the JUN45 effective interaction~\cite{honma09,kaneko14} including isospin non-conserving terms. In addition, from the present data the mirror energy differences can be determined for the second $2^+_2$ state, $\text{MED}(2^+_2) = -122(11)$~keV, as well as for the $4^+_2$ candidate $\text{MED}(4^+_2) = -28(25)$~keV. \nuc{70}{Kr} is the heaviest $T_z = -1$ nucleus where the $2^+_2$ state and therefore the mirror energy difference is known. The next heaviest case is \nuc{58}{Zn}~\cite{langer14}. In all known cases the mirror energy difference for the $2^+_2$ state is negative and, with the exception of the \nuc{34}{Ar}-\nuc{34}{S} system, larger in magnitude than for the $2^+_1$ state. For the $4^+_2$ state the only other known case is $A=26$.

Besides the shell model calculations of the mirror energy differences various beyond mean field approaches have been used to calculate \nuc{70}{Kr} and neighboring nuclei~\cite{petrovici03,girod09,hinohara10,rodriguez14}. Calculations based on the excited VAMPIR approach predict a strong prolate-oblate mixing with about equal contributions in \nuc{70}{Se} to the yrast states and a slight prolate dominance in \nuc{70}{Kr}~\cite{petrovici15}. Furthermore, an excited band with more oblate configurations is also predicted~\cite{petrovici17}.
Recently, collective properties of nuclei along the whole Kr isotopic chain have been calculated using the symmetry conserving configuration mixing (SCCM) approach with the Gogny D1S interaction~\cite{rodriguez14}. These calculations show the importance of the triaxial degree of freedom in the light Kr isotopes. The calculated level schemes for \nuc{70}{Kr} and \nuc{70}{Se} are compared to experimental ones in Fig.~\ref{fig:level}. Overall good agreement is observed, the calculated level energies are in general lower than the experimental ones, similar to other cases where these calculations overestimate the collectivity. For \nuc{70}{Kr} and \nuc{70}{Se} they predict an oblate deformed ground state band with $\beta_\text{obl}\sim0.35$. A second, strongly prolate deformed band is predicted in both nuclei with $\beta_\text{pro}\sim0.55$. While low-lying excited $0^+$ states have not been observed in the present experiment -- the $0^+_2$ state at 2010~keV in \nuc{70}{Se} is not a candidate for the band-head -- the excitation energies of the $2^+_2$ and $4^+_2$ states match reasonably well the calculations.
Lastly, calculations using a five-dimensional collective Hamiltonian (CHFB-5DCH)~\cite{girod09,delaroche10} are compared to the experimental results. These global calculations capture the collective features across the nuclear chart. For \nuc{70}{Kr} and \nuc{70}{Se} they also predict an oblate ground state band, with prolate deformation in the yrare states. The excitation energies of the $2^+_1$ states are better reproduced than with the SCCM calculations, but the $2^+_2$ state is predicted too high with respect to the $2^+_1$ and $4^+_1$ states.
In general, all theoretical approaches describe the excitation energy spectrum reasonably well. They all predict shape coexistence with an oblate ground state band, except for the VAMPIR calculations which has prolate dominant configurations in the ground state band. 

\begin{figure}[h]
\centering
\includegraphics[width=\columnwidth]{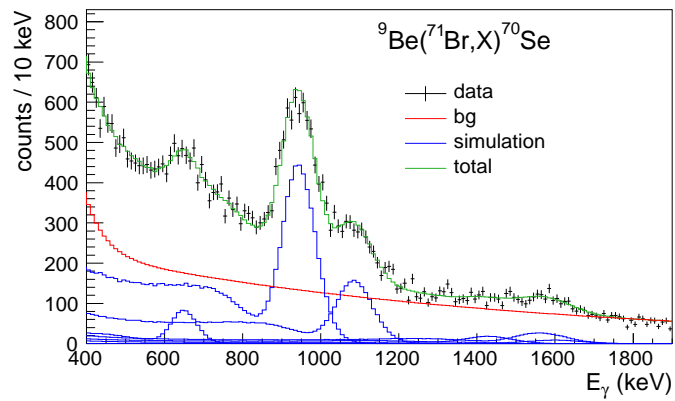}
\caption{$\gamma$-ray energy spectra for \nuc{70}{Se} populated in one-proton removal reactions from a \nuc{71}{Br} beam. Measured $\gamma$-ray energies have been corrected for the Doppler-shift using the position of DALI2 crystals with respect to the target. The data are superimposed with the result of a likelihood fit of a continuous background and GEANT4 simulations of the DALI2 response function.}
\label{fig:segamma}
\end{figure}

Fig.~\ref{fig:segamma} shows the $\gamma$-ray energy spectrum for \nuc{70}{Se} populated by one-proton knockout from \nuc{71}{Br} measured in the same experiment. This reaction mirrors the one-neutron knockout reaction shown in Fig.~\ref{fig:gamma}. 
The broad structure around 1400 - 1600~keV in the \nuc{70}{Se} spectrum shown in Fig.~\ref{fig:segamma} probably contains several transitions. The 1483~keV transition from the known $4^+_2$ state, the 1574~keV transitions from the $3^-$ state, a 1609~keV transition from a $(4^+_3)$ candidate, and the 1600~keV transition from the $2^+_2$ state to the ground state. The intensity of the latter has been fixed in the fit using the known branching ratio of 25(5)~\% compared to the 665~keV transition. The uncertainty of the other intensities is large due to the limited resolution of DALI2.

The inclusive cross sections for the two reactions have been determined from the number of measured particles in the BigRIPS and ZeroDegree spectrometers, the transmission through the spectrometers, and the detection efficiency of the beam-line detectors. In the case of \nuc{70}{Se} the transmission of the reaction products through the ZeroDegree spectrometer was limited by the acceptance and therefore the extrapolation of the cross section beyond the acceptance limit resulted in a larger systematic uncertainty. For the proton removal from \nuc{71}{Br} the cross section amounts to $\sigma^{-1p}(^{70}\text{Se}) = 100(4)(11)$~mb, while the neutron removal reaction from \nuc{71}{Kr} has a cross section of $\sigma^{-1n}(^{70}\text{Kr}) = 24.5(9)(9)$~mb. This large asymmetry of the reaction cross sections is due to the reaction dynamics as well as to nuclear structure effects. One would expect a ratio of $\sigma^{-1n}(^{70}\text{Kr})/\sigma^{-1p}(^{70}\text{Se}) \sim 0.4$ arising from the asymmetry in binding energies~\cite{tostevin14} and the difference in single-particle cross sections. The single-particle cross section for the removal of a nucleon depends on the binding energy of the removed nucleon. Calculations using the formalism of~\cite{tostevin14} show that for the relevant orbitals $1f_{7/2}$, $2p_{3/2}$, $1f_{5/2}$, $2p_{1/2}$, and $1g_{9/2}$, for the same excitation energy the single-particle cross sections for the removal of a neutron from \nuc{71}{Kr} is about $70 - 80$~\% of the cross section value of the analogue removal of a proton from \nuc{71}{Br}. In knockout reactions a reduction of the spectroscopic strength compared to shell model calculations has been observed~\cite{tostevin14,gade08}. This reduction depends strongly on the asymmetry in binding energy. These reduction factors amount to $R_n\approx0.4$ for neutron removal and $R_p\approx0.8$ for proton removal of the nuclei around $N=Z$. Assuming isospin symmetry and thus identical structure of the beams \nuc{71}{Kr} and \nuc{71}{Br}, and the reaction products \nuc{70}{Kr} and \nuc{70}{Se}, the cross section for \nuc{70}{Kr} should be a factor of 0.4 lower than the one for \nuc{70}{Se}. 

\begin{figure}[h]
\centering
\includegraphics[width=\columnwidth]{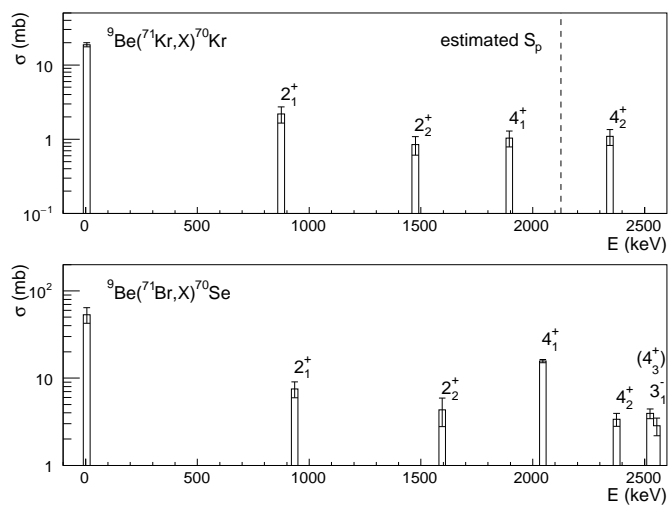}
\caption{Measured final state exclusive cross sections for the population of individual states in \nuc{70}{Se} and \nuc{70}{Kr} through the one-proton and one-neutron removal reactions, respectively. Also shown is the estimated proton separation energy $S_p(^{70}\text{Kr}) = 2.1$~MeV.}
\label{fig:crosssections}
\end{figure}

The proton separation energy of \nuc{70}{Kr} is estimated at $S_p(^{70}\text{Kr}) = 2.1$~MeV from systematics~\cite{audi12}. As shown in Table~\ref{tab:intensities} and Fig.~\ref{fig:crosssections}, the population of highly excited states in \nuc{70}{Se} is not negligible. In \nuc{70}{Kr} the corresponding states could be unbound.
Structural differences between the two reactions also include potentially different ground state spins and isomeric contamination for the $A=71$ beams.

From the measured $\gamma$-ray yields final state exclusive cross sections for the population of individual states in \nuc{70}{Se} and \nuc{70}{Kr} have been determined. These are compared in Fig.~\ref{fig:crosssections} and Table~\ref{tab:intensities}.
The relative population of the $0^+_1$, $2^+_1$ and $2^+_2$ states is very similar in the two cases, but the \nuc{9}{Be}(\nuc{71}{Br},X)\nuc{70}{Se} reaction favors the population of higher spin states of $4^+$ and $3^-$. This can be due to further unobserved feeding from higher lying states, which in the case of \nuc{70}{Kr} are unbound, and therefore do not reach the ZeroDegree focal plane as \nuc{70}{Kr}. In \nuc{70}{Se} the $4^+_1$ and $0^+_2$ states are only separated by 29~keV. In the fit shown in Fig.~\ref{fig:segamma} only the $4^+_1\rightarrow 2^+_1$ transition was considered. However, the extracted intensity could be a mixture of $4^+_1$ and $0^+_2$ population. The ground state spin of \nuc{71}{Kr} is unknown but expected to be $(5/2^-)$~\cite{fisher05}, the ground state of \nuc{71}{Br} is tentatively assigned as $(5/2^-)$ with an excited $(1/2^-)$ state at 10~keV. Consequently the ground state of \nuc{71}{Kr} could be either $(5/2^-)$ or $(1/2^-)$ and an isomeric state with the other spin is expected at low excitation energy. We have calculated theoretical cross sections for both reactions based on shell model spectroscopic factors using the JUN45 effective interaction~\cite{honma09}. These show that the population of the ground state as well as the $4^+_1$ state is largest when the initial state of the projectile nucleus is assumed to be the $5/2^-$ state. While in both cases the beam could potentially in an isomeric state the main component seems to be the $5/2^-$ state also in the case of \nuc{71}{Kr}.

In summary, the spectroscopy of \nuc{70}{Kr} has been clarified and significantly extended. Five excited states were placed in the level scheme based on a $\gamma-\gamma$ coincidence analysis. Comparison to the mirror nucleus \nuc{70}{Se}, as well as the selectivity of the different reaction channels, allowed for tentative spin and parity assignments. The mirror energy differences obtained for the $2^+_2$ and $4^+_2$ states are the heaviest known cases for $T=1$ nuclei. Theoretical calculations reproduce the existence of a second band with different deformation, and thus shape coexistence in \nuc{70}{Kr}. Further investigations of the quadrupole collectivity through Coulomb excitation will pin down the nature of the deformation in the future.

We would like to thank the RIKEN accelerator and BigRIPS teams for providing the high intensity beams and T. R. Rodr\'iguez for providing us with the calculations for \nuc{70}{Se}. 
This work has been supported by UK STFC under grant numbers ST/L005727/1 and ST/P003885/1, the Spanish Ministerio de Econom\'ia y Competitividad under grants FPA2011-24553 and FPA2014-52823-C2-1-P and the Program Severo Ochoa (SEV-2014-0398).
AO thanks the support from the European Research Council through the ERC Grant No. MINOS-258567.

\bibliographystyle{elsarticle-num-names}
\bibliography{draft}

\end{document}